\documentclass[12pt]{rspublic}
\usepackage[
           margin=1.5cm,
           a4paper,
           includehead
   ]{geometry}
\usepackage{amsfonts}
\usepackage{natbib}
\usepackage{graphicx}
\newcommand{\beq}{\begin{eqnarray}}
\newcommand{\eeq}{\end{eqnarray}}
\newcommand{\diff}{\mbox{d}}

\begin{document}
\title{Strongly Interacting Dynamics beyond the Standard Model on a Spacetime Lattice}
\author{Biagio Lucini}
\affiliation{School of Physical Sciences, Swansea University, Singleton Park, Swansea SA2 8PP, UK}
\date{}
\maketitle
\begin{abstract}{Mechanism of Electroweak Symmetry Breaking, Strongly Interacting BSM dynamics, Lattice Gauge Theory}
Strong theoretical arguments suggest that the Higgs sector of the Standard Model of the Electroweak interactions is an effective low-energy theory, with a more fundamental theory that is expected to emerge at an energy scale of the order of the TeV. One possibility is that the more fundamental theory be strongly interacting and the Higgs sector be given by the low-energy dynamics of the underlying theory. We review recent works aimed to determining observable quantities by numerical simulations of strongly interacting theories proposed in the literature for explaining the Electroweak symmetry breaking mechanism. These investigations are based on Monte Carlo simulations of the theory formulated on a spacetime lattice. We focus on the so-called Minimal Walking Technicolour scenario, a SU(2) gauge theory with two flavours of fermions in the adjoint representation. The emerging picture is that this theory has an infrared fixed point that dominates the large distance physics. We shall discuss the first numerical determinations of quantities of phenomenological interest for this theory and analyse future directions of quantitative studies of strongly interacting beyond the Standard Model theories with Lattice techniques. In particular, we report on a finite size scaling determination of the chiral condensate anomalous dimension $\gamma$, for which we find $0.05 \le \gamma \le 0.25$. 
\end{abstract}
\section{Introduction}
Despite the experimental success of the Standard Model in its current formulation (see e.g. the results for the precision tests reported in~\cite{Amsler:2008zzb}), its Higgs sector (which is responsible for giving mass to the $W$ and $Z$ bosons and to fermionic matter) is firmly believed to be an effective theory. The main theoretical arguments for this are: (1) there are no other fundamental scalars in nature, and in other systems where the Higgs mechanism is at work (e.g. superconductors) the Higgs boson is never a fundamental particle; (2) the Mexican hat shaped potential is put in the Lagrangian by hand; (3) the renormalised Higgs mass is 16 orders of magnitude smaller than the Lagrangian mass ({\em hierarchy problem}). All these aspects of the theory (especially the fine tuning of the mass) are seen as unnatural for a fundamental theory. Hence, it must exist a more fundamental theory at an energy scale above the natural cut-off of the Standard Model (1 TeV) of which the Standard Model itself is the low-energy manifestation. This theory must contain a mechanism of mass generation for the massive gauge bosons and for the fermionic matter and at the same time fulfil the severe experimental bounds of precision measurements in the Electroweak sector. 

Various scenarios have been conjectured to extend the Standard Model above 1 TeV (see e.g.~\cite{Accomando:2006ga} for an overview). Among them, strongly interacting dynamics beyond the Standard Model is a framework in which massive gauge bosons get their mass from an interaction with fermions in a gauge theory that contains the Electroweak gauge group SU(2)$_L$ $\otimes$ U(1)$_Y$ ({\em Technicolour}, \cite{Weinberg:1975gm} and \cite{Susskind:1978ms}): the strong interactions break chiral symmetry, generating a chiral condensate (which plays the role of the Higgs condensate in the Standard Model) and the associated massless Goldstone particles, which are then reabsorbed by some gauge bosons to give rise to a longitudinal component for them. Standard Model fermions get mass from a weakly coupled interaction at higher scale ({\em Extended Technicolour}, see~\cite{Eichten:1979ah} and \cite{Dimopoulos:1979es}). While still characterised by a large coupling to give rise to phenomena like confinement and chiral symmetry breaking, in order to conciliate the experimental values of masses of quarks with experimentally unobserved flavour changing neutral currents, the beyond the Standard Model strong interaction can not be a simple replica of the Standard Model strong sector. An appealing possibility in principle compatible with observations is that the theory have a large anomalous dimension for the chiral condensate (\cite{Holdom:1984sk}; \cite{Yamawaki:1985zg}; \cite{Appelquist:1986an}). This is realised if for the particular choice of the number of flavours and of the number of colours the theory has or is near to an infrared fixed point. This property of infrared conformality (or near-conformality) could also solve the tension with precision Electroweak measurements pointed out in~\cite{Peskin:1991sw}. For a review of the subject, see e.g.~\cite{Hill:2002ap}.  

In order to be near an infrared conformal point, a SU($N$) gauge theory with fermions in the fundamental representation must have a number of fermions close to 4$N$. This large number of flavours would result in a plethora of still to be discovered particles. Recently, it has been pointed out in~\cite{Sannino:2004qp} that to reduce the number of fermion flavours for the confining theory to be near an infrared fixed point, fermionic matter can be in a two-index (symmetric or adjoint) representation. Since then, the extent of the conformal phase for these theories has been computed using various techniques (see e.g.~\cite{Dietrich:2006cm} or more recently~\cite{Armoni:2009jn}, \cite{Sannino:2009me} and~\cite{Poppitz:2009tw}). Although providing estimates for the extent of the conformal window in broad agreement, all these calculations are based on uncontrolled expansions and/or educated guesses. Since this problem is crucial for the viability of Technicolour as a mechanism of Electroweak symmetry breaking, an investigation from first principles of the extent of the conformal window and of observables of phenomenological interest at his lower end is needed. In this article, we shall review calculations performed in this spirit for an SU(2) gauge theory with two flavours of Dirac fermions in the adjoint representation. Since this is the (vectorial) theory with the minimal content of particles that could be compatible with phenomenological requirements for a strongly coupled extension of the Standard Model, it is referred to as {\em Minimal Walking Technicolour}. The framework used for the calculations discussed in this paper is Lattice Gauge Theory. An effort is made to keep the material accessible to a non-specialised audience, which will necessarily result in the omission of important technical details. The reader interested in more technical discussions or in lattice studies of models different from the one discussed in this paper is referred to the specialised literature. The recent activity in this and in related sectors has been reviewed by \cite{Fleming:2008gy} and \cite{Pallante}.
\section{Confining vs. Infrared conformal behaviour}
The prototype of a strongly interacting theory is Quantum Chromodynamics (QCD). This is a theory that couples eight gauge bosons to six flavours of fermions in the fundamental representation. The gauge group is SU(3). One of the features of this theory is that at low energies the coupling is strong. Because of that, the resulting phenomenology is determined by confinement and chiral symmetry breaking. Confinement is the statement that the fundamental matter fields, the quarks, do not exist as free asymptotic states. In fact, the states that are experimentally observed transform trivially under SU(3). This can be understood in terms of a linearly increasing potential that binds a quark to an antiquark: $V(r) = \sigma r$, with $\sigma$ the string tension.

In the limit in which the Lagrangian mass is set to zero, chiral symmetry is spontaneously broken: a non-zero quark condensate forms and massless Nambu-Goldstone bosons associated with the spontaneous symmetry breaking appear. These bosons are pseudoscalar particles. In fact, in real world QCD the lightest quarks do have a small mass that breaks explicitly chiral symmetry, but the breaking is soft: pseudoscalar bosons are still the lightest states of the theory, with the mass of the vectors around a factor of four larger and the axial states are higher in mass than their vectorial partners.

The observation that the chiral condensate breaks the Electroweak gauge group $SU(2)_L \otimes U(1)_Y$ lead to the original formulation of Technicolour as a mechanism of Electroweak symmetry breaking. However, it was immediately manifest that a theory that simply mimics QCD would be at odd with experiments. The problem of a QCD-like theory can be traced back to the existence of only one energy scale. The phenomena of chiral symmetry breaking and confinement arise both at a dynamically generated scale of energy called $\Lambda_{QCD}$. For energies lower than $\Lambda_{QCD}$ the theory is non-perturbative, while above $\Lambda_{QCD}$ the theory is perturbative. There is no reason a priori why the scale controlling the onset of asymptotic freedom should be the same as the ones controlling the onset of confinement and chiral symmetry breaking\footnote{Moreover, unlike in QCD, the latter two scales could be different.}. If this happens and in a regime of energies intermediate between the two scales the system displays near-conformal infrared dynamics, the theory could be a candidate for a strongly interacting extension of the Standard Model. The key point is that a large anomalous dimension for the chiral condensate should be generated by the interactions in order for the theory to be consistent with experimental results.  

Excluding the pseudoscalar channel, in confining theories with chiral symmetry breaking correlation functions of operators decay as an exponential governed by the lowest mass in the channel with the quantum numbers of the observable considered. On the contrary, in a conformal theory there is no mass gap in any channel and correlators have a power law behaviour. In this case, correlators at separation $r$ do not behave as $r^{-D}$, where $D$ is the scaling dimension of the operator given by dimensional analysis, but as $r^{-D-\eta}$. $\eta$, which is determined by the interactions, is called the anomalous dimension. An infrared conformal theory is a theory that appears conformal only at large distances. Such a theory is similar to a statistical system at criticality. For near-conformal behaviour, the system behaves like an infrared conformal theory in a certain interval of distances, but at some higher distance becomes confining and chiral symmetry breaking. Still using a Statistical Mechanics language, near-conformality is a cross-over phenomenon. Hence, a good understanding of infrared conformal theories can be used as a starting point for a study of nearly-conformal ones.

As in Statistical Mechanics it is useful to consider a relevant interaction that drives the system outside the critical surface and to study how the critical point is reached when the strength of this interaction is sent to zero, it is interesting to consider soft breaking of the conformal symmetry. One possibility is to add a small mass term $m$ to the Lagrangian. In this case, a mass gap $M \propto m^{1/(1+\gamma)}$ is generated, where $\gamma$ is the anomalous dimension of the chiral condensate (or equivalently $- \gamma$ is the anomalous dimension of the mass). In the more conventional scenario, the exponent $\gamma$ governs the behaviour of all infrared quantities with naive mass dimension one. Moreover, at energy scales $E \ll M$ an effective Yang-Mills SU($N$) gauge theory might emerge. An illustration of this latter feature has been provided by~\cite{Miransky:1998dh}: in a system that admits a weakly-coupled infrared fixed point, the author has shown that an energy scale $\Lambda \ll M$ is generated and the long distance physics is given by glueball states, while mesons decouple from the dynamics like in QCD in the limit in which $m \to \infty$. This phenomenon can be described as dynamical quenching of the theory: fermion loops do not contribute to the large-distance physics and can then be neglected at low energy. An interesting feature of this picture is that $\Lambda/M$ is constant (at least in a regime in which perturbation theory can be trusted): hence, it does not matter how small $m$ is, the theory will always dynamically quench at some low enough energy scale, and the large-distance behaviour will always be that of quenched QCD. The dynamical generation of a scale proportional to $M$ is called locking. 

\begin{figure}[t]
\begin{tabular}{cc}
\includegraphics*[scale=0.45]{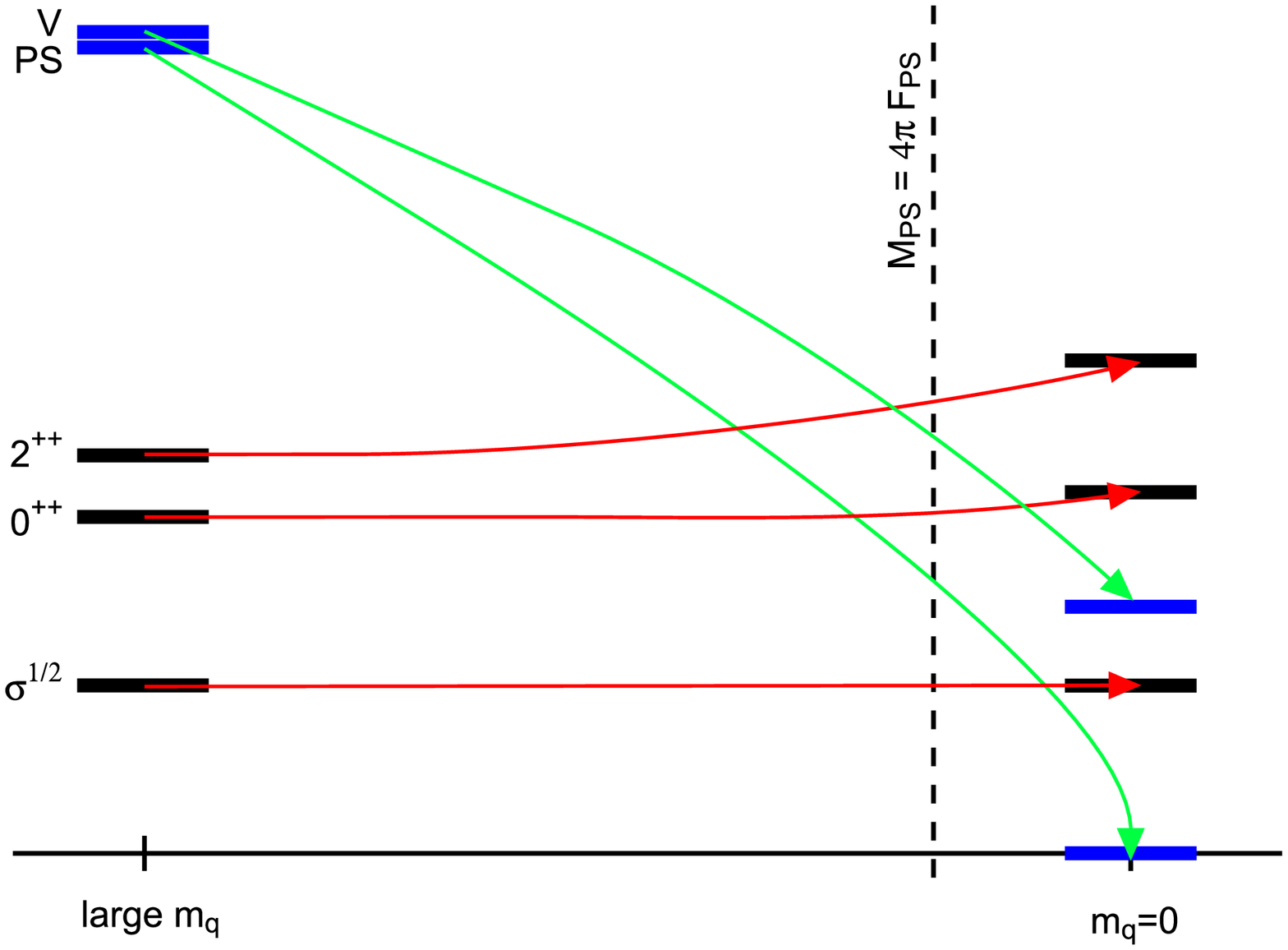}& 
\includegraphics*[scale=0.45]{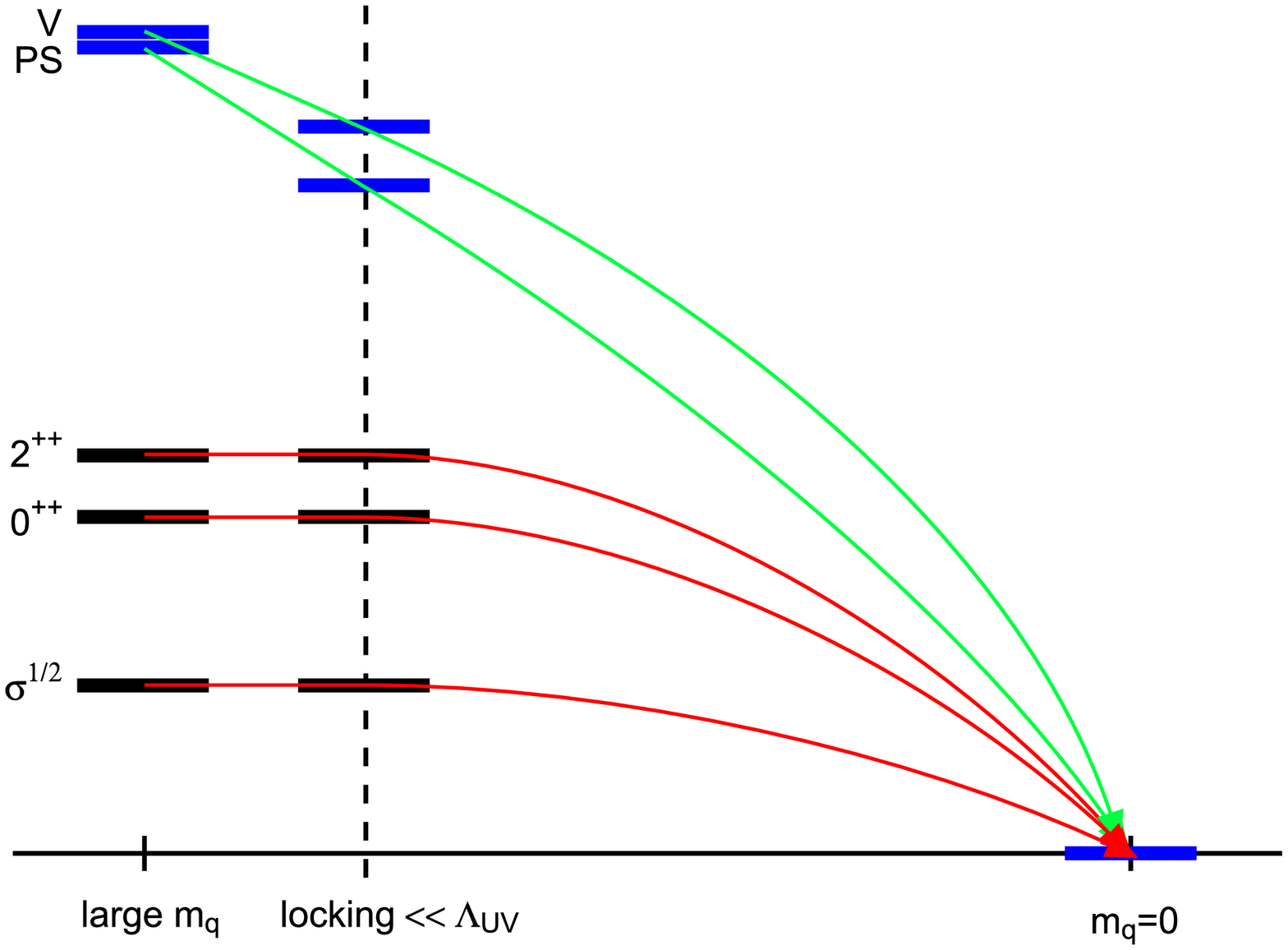}\\
(a) & (b)
\end{tabular}
\caption{Evolution of the spectrum as a function of $m$ for (a) a confining and
chiral symmetry breaking theory and (b) an IR conformal theory with dynamical
quenching. Artwork by A. Patella.}
\label{fig:1}
\end{figure}

Dynamical quenching, suppression of the Yang-Mills scale and locking (which we shall refer to as the Miransky scenario) make the spectrum of an infrared (IR) conformal field theory softly broken with a mass term significantly different from that of QCD, but strictly speaking the existence of those phenomena can only be proven at weak coupling. However, some or all the features of the Miransky scenario could in principle survive near a more strongly coupled IR fixed point. The surviving features (if any) could be powerful enough to allow us to unambiguously assess if a SU($N$) gauge theory with a small mass is confining or IR conformal. Let us assume for instance that only locking takes place in a more general case for IR conformal theories. In a SU($N$) gauge theory with fermionic matter and a Lagrangian mass $m \gg \Lambda_{UV}$, where $\Lambda_{UV}$ is the largest dynamically generated scale (the scale that controls the onset of asymptotic freedom), in both the IR conformal and the confining case the theory can be described by an heavy-quark effective field theory, which features a nearly-degenerate meson spectrum much heavier than glueball states. As the mass is lowered, the character of the theory in the IR starts to emerge: in a confining and chiral symmetry breaking theory the pseudoscalar meson becomes progressively lighter and eventually massless, while the vector meson has a non-zero mass in the chiral limit that is smaller than the typical glueball mass; if the theory is IR conformal, at some mass $\tilde{m} \ll \Lambda_{UV}$ mass ratios in the spectrum become approximately constant in $m$, with the whole spectrum collapsing to zero in the massless limit. The features of the spectrum at energies less than $\tilde{m}$ (e.g. whether the pseudoscalar is the lightest state) are determined by the features at $\tilde{m}$, where the snapshot was taken. In other words, dynamical quenching (and the associated suppression of the Yang-Mills scale) is not the only possible scenario in the IR conformal theory. This argument seems to show that locking is the more fundamental property of the Miransky scenario. The IR conformal vs. the confining and chiral symmetry breaking behaviour as the mass is varied is illustrated pictorially in Fig.~\ref{fig:1} for the case in which the IR conformal theory is characterised by dynamical quenching. Finally, in a nearly-IR conformal theory the spectrum will freeze in a range of energies between $\Lambda_{IR}$ and $\tilde{m}$, where $\Lambda_{IR}$ is another dynamically generated scale. Below $\Lambda_{IR}$ the signature would still be that of a confining and chiral symmetry breaking theory.

It is worth stressing once again that while the scaling of all masses as $m^{1/(1+\gamma)}$ comes from general scaling arguments, the survival of a Miransky-like or related scenario beyond perturbation theory should be investigated from first principles. {\em Ab-initio} calculations are also needed to determine quantitative features of (nearly-) IR conformal theories, like e.g. whether $\gamma$ is of order one or more, as Technicolour models prefer and some semi-analytical arguments advocate.
\section{Gauge Theories on a Spacetime Lattice}
Weakly coupled gauge theories can be defined by perturbing around the free theory vacuum. For a consistent definition of strongly coupled gauge theories this procedure is unreliable. An interesting possibility is to use the path integral formulation regularised on a spacetime lattice. On a discrete lattice of spacing $a$ and linear extension $L$ the Wightman functions of the theory  are well defined. The continuum theory is recovered performing the limits for $L \to \infty$ and for $a \to 0$ of observables in the discretised theory. This scheme of calculations does not rely on any approximation, hence, it is a scheme that can provide results from first principles. In addition, it has the virtue of not spoiling the gauge-invariance of the theory. If analytical continuation to imaginary time is used, the path integral becomes formally identical to the partition function of a system in Statistical Mechanics, and numerical methods based on Monte Carlo simulations can be used to compute observables as averages over the most important configurations of the system with an incertitude that is statistically under control in terms of the number of the generated configurations. We consider lattice geometry $N_t \times N_s^3$, where the integer $N_t$ measures the extension of the lattice in the temporal direction and $N_s$ is the spatial extension. 

The lattice action $S$ of a SU($N$) gauge theory with fermionic matter can be written as the sum of the gauge term $S_g$ and the fermion contribution $S_f$. Various choices can be performed for $S_g$ and $S_f$ that reproduce the correct continuum behaviour in the $a \to 0$ limit. The simplest choice for $S_g$ is the Wilson action in the fundamental representation
\beq
\label{wilson}
  S_g = \beta\sum_{i,\mu > \nu}\left( 1 - \frac{1}{N} {\cal R}\rm{e~}\rm{Tr}(U_{\mu \nu}(i))
  \right) \ , 
\eeq
where $\beta = 2N/g^2$ and $g$ is the coupling. The plaquette $U_{\mu \nu}(i) = U_{\mu}(i) U_{\nu}(i+\hat{\mu}) U_{\mu}^{\dag}(i+\hat{\nu})U_{\nu}^{\dag}(i)$  is the parallel transport along the smallest closed contour ({\em plaquette}) in $i$ identified by the directions $\hat{\mu}$ and $\hat{\nu}$ of the link variables $U_{\mu}(i)$ defined on lattice links stemming from $i$ to $i + \hat{\mu}$. $i = (n_1, n_2, n_3, n_4)$ is a lattice point labelled by four integer coordinates. When a link is followed in a negative direction, the Hermitian conjugate of the link variable has to be taken. ${\cal R}\rm{e~}\rm{Tr}(U_{\mu \nu}(i))$ in Eq.~(\ref{wilson}) indicates that the real part of the trace of the plaquette has to be taken; the sum is over all lattice points and over all directions. The action $S_g$ is manifestly gauge invariant: when performing a gauge transformation $\tilde{U}_{\mu}(i) = G^{\dag} (i) U_{\mu}(i) G(i + \hat{\mu})$, with $G(i)$ elements of SU($N$), $S_g$ is not affected. In terms of the continuum gauge field $A_{\mu}$
\beq
U_{\mu}(i) = Pexp\left( ig \int_i^{i+a\hat{\mu}} A_{\mu}(x)\diff x \right) \ , 
\eeq
with the integral performed on the link connecting $i$ and $i+a\hat{\mu}$. It is easy to show from this definition that the Wilson action reproduces the continuous gauge action when $a \to 0$.

Since the fermion contribution is quadratic in the fermion fields, in the path integral fermion variables can be integrated explicitly. The result is the determinant of the quadratic form $M$ in the fermion action to the power of $N_f$. A naive approach for the discretisation of fermion fields on the lattice determines the appearance of extra fermion flavours, due to the explicit breaking of the Lorentz symmetry. Various solutions have been proposed for this problem. In this work, we use the Wilson proposal, in which an irrelevant operator that makes the extra flavours infinitely heavy in the continuum limit is added to the lattice action. In the Wilson formulation, the fermionic quadratic form ({\em Dirac operator}) reads
\beq
  M_{\alpha \beta}(ij) = (m+ 4r) \delta_{ij}
  \delta_{\alpha \beta} - \frac{1}{2} \sum_{\mu} \left[\left(r \mathbb{I} - \gamma_{\mu}\right)_{\alpha
      \beta}U^A_{\mu}(i) \delta_{i,j+\hat{\mu}} + \left(r \mathbb{I} +
      \gamma_{\mu}\right)_{\alpha \beta} U_{\mu}^{A \dag}(j)\delta_{i,i-\hat{\mu}}
  \right] \ ,
  \eeq
where $r$ multiplies the irrelevant operator added to the action (conventionally, $r = 1$ is taken) and the indices $\alpha$ and $\beta$ are spinorial. $m$ is the bare fermion mass in units of the cut-off $a$, and $U^A_{\mu}(i)$ is the link in the adjoint representation. The main drawback of the Wilson discretisation of fermions is the explicit breaking of the chiral symmetry due to the non-zero $r$. Since there is no symmetry to protect the mass from an additive renormalisation, the value of the mass (or equivalent, of the hopping parameter $\kappa = 1/(2(m+ 4 r))$ at which the system is chiral is determined by quantum effects and needs to be computed. A more technical presentation of the methods and the observables studied for the Minimal Walking Technicolour case is contained in \cite{DelDebbio:2008zf}, \cite{DelDebbio:2009fd} and \cite{Pica:2009hc}.  
\section*{The emerging picture}
Lattice Gauge Theory can build upon a legacy of more than 30 years. The simultaneous evolution of computational techniques and computational power (most of the computations are performed on state of the art supercomputers, sometimes designed and developed by lattice gauge theorists for their simulations) has generated an increasing level of activity. Currently most parameters in the strong sector of the Standard Model can be determined with a precision of the order of a few percents. In beyond the Standard Model calculations, this level of precision is not required. However, simulating theories beyond the Standard Model is more difficult, due to the lack of experimental guidance and adequate analytical understanding. In fact, the lattice naturally introduces systematic errors related to the ultraviolet cut-off $a$ and the infrared cut-off $L$, the latter being relevant in our context since it explicitly breaks conformal symmetry. In addition, for technical reasons a fermion mass term needs to be explicitly added, which mandates an extrapolation to the chiral limit. Taking correctly (and in the right order) the limits $L \to \infty$, $a \to 0$ and $m \to 0$ is crucial for capturing the right features of the continuum massless theory formulated in an infinite volume. For instance, if the lattice spacing is reasonably small but the lattice size is not large enough, the resulting spectrum can be significantly distorted (in practise, a gauge theory will always look conformal on a small box, since in this setup only the short-distance regime can be explored, and this is perturbative). Conversely, if the volume is large but the lattice is too coarse, the lattice theory can be in a phase that is not the phase of the continuum theory (in the lattice strong coupling regime, a SU($N$) gauge theory is always confining). Hence, an initial investigation of the phase structure of the theory on the lattice is mandatory. For those reasons, Monte Carlo simulations need to be performed in the phase connected with the continuum and in a regime in which the inverse lattice spacing is much larger and the inverse lattice size is much smaller than the mass of the lightest state of the theory. While the infinite size and zero lattice spacing regimes look easier to reach in principle, it is more difficult to define a small mass regime for a theory whose chiral behaviour is not fully understood.

The study of Minimal Walking Technicolour with Lattice techniques was started very recently (\cite{Catterall:2007yx}; \cite{DelDebbio:2008zf}), with the phase structure of the theory systematically explored in \cite{Catterall:2008qk} and \cite{Hietanen:2008mr}. One striking feature that emerged from these early works is the near-degeneracy of the pseudoscalar and vector mesons. This near-degeneracy could be a signature of an IR conformal behaviour of the theory. However, these investigations lacked control over the continuum, the chiral and the infinite volume extrapolations. Hence, from their results it can not be excluded that the observed behaviour would be irrelevant for the massless continuum theory. In the last year, studies have appeared (\cite{Pica:2009hc}; \cite{Catterall:2009sb}) where progress is made to control the chiral and the infinite volume extrapolations. The nearly-degeneracy in the meson spectrum is confirmed at smaller masses, and seems to be a feature of the chiral limit. Moreover, the meson decay constant, $f_{\pi}$, is studied. This quantity is observed to have large finite size corrections at small $m$, and seems to go to zero in the chiral and infinite size limits, compatibly with the existence of an IR fixed point. However, a fundamental question still stands: how can we be sure that we are observing a theory in the small mass limit in a large volume and not a massive theory in a small volume? Notice that at this stage there is no physical quantity that can be unambiguously used to establish whether the Lagrangian mass is small or to quantify the size of the lattice in physical units.

\begin{figure}[t]
\vspace{0.25cm}
\begin{minipage}[b]{0.5\linewidth}
\centering
\includegraphics[scale=0.35]{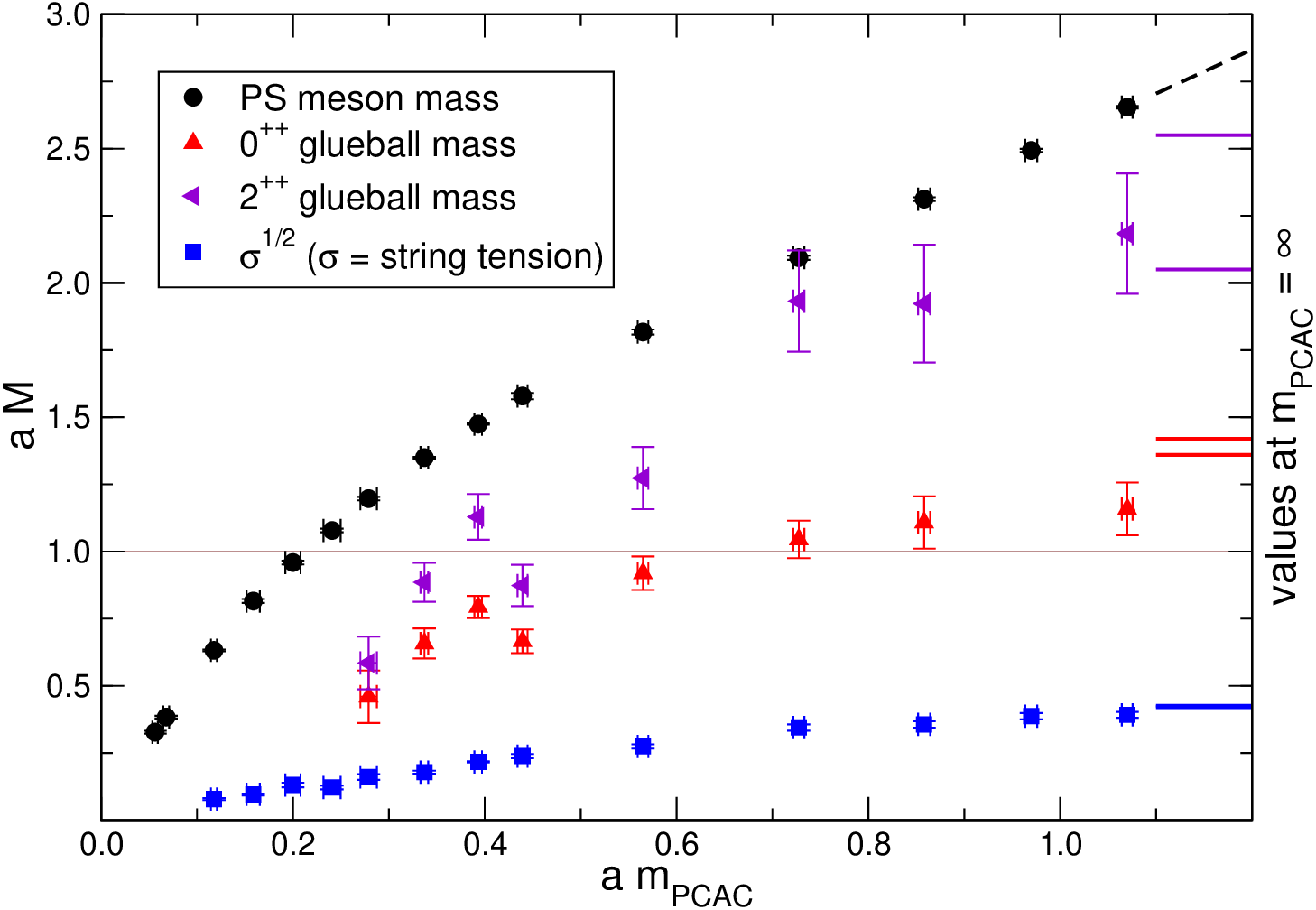}
\caption{The mass spectrum of Minimal Walking Technicolour at $\beta = 2.2$ as a function of the fermion mass $m_{PCAC}$ in units of the lattice spacing $a$.}
\label{fig:2}
\end{minipage}
\hspace{0.1cm}
\begin{minipage}[b]{0.5\linewidth}
\centering
\includegraphics[scale=0.35]{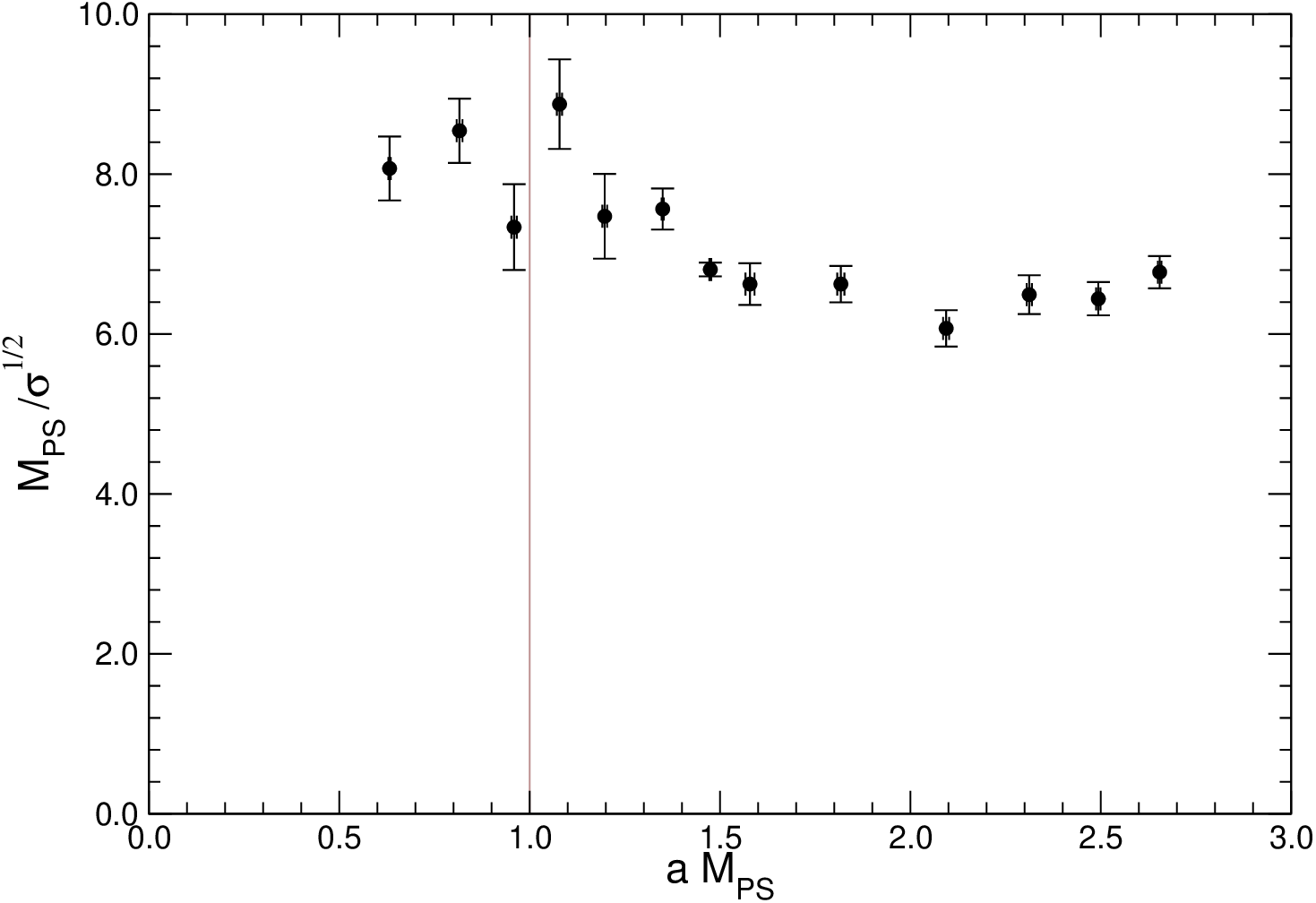}
\caption{The ratio $M_{PS}/\sqrt{\sigma}$ shows a mild dependence on the the pseudoscalar mass $M_{PS}$, plotted in units of the lattice spacing $a$.}
\label{fig:3}
\end{minipage}
\end{figure}
Progress in understanding the significance of lattice results for the continuum massless theory were made in~\cite{DelDebbio:2009fd}. The important observation in this work is that the ambiguity can be resolved by looking at quantities in the gluonic sector in addition to the fermionic quantities already studied in the literature. In fact, there is a well defined hierarchy in the spectrum, with the string tension being the smallest scale and the glueball spectrum being much lighter than the meson spectrum (see Fig.~\ref{fig:2}, where the vector meson is not displayed because on the scale of the figure it looks degenerate with the pseudoscalar). Together with the associated locking of the scale (Fig.~\ref{fig:3}), this is a strong evidence for the Miransky scenario with suppression of the scale and dynamical quenching of the spectrum (compare the lattice data in Fig.~\ref{fig:2} with the right panel of Fig.~\ref{fig:1}). A comparison with the spectrum of the Yang-Mills theory after having adjusted the bare parameters in such a way that relevant physical quantities are tuned in the two theories shows that the observed spectrum is in fact quenched and the ratio between the mass of the vector and of the pseudoscalar is $M_{V}/M_{PS} \simeq 1.04$ for small Lagrangian masses. A useful quantity to use to characterise the fermion mass is the mass extracted from the axial Ward identity, $m_{PCAC}$, defined as
\begin{eqnarray}
  \label{eq:pcacplateau}
  m_{PCAC}= \lim_{t \to \infty} \frac{1}{4} \frac{C_{\gamma_0 \gamma_5,\gamma_5} (t+1)
- C_{\gamma_0 \gamma_5,\gamma_5} (t-1) } {C_{\gamma_5,\gamma_5} (t)} 
\end{eqnarray}
(see e.g. Sect.~5 in~\cite{DelDebbio:2008zf} for details).

Independent evidence for the existence of an IR fixed point comes from studies of the running of the coupling: \cite{Hietanen:2009az} and \cite{Bursa:2009we} show that at large distance the gauge coupling of the theory flows to a fixed value $g^{\star}$. Although the fixed point value depends on the choice of a scheme of regularisation (the Schr\"odinger functional has been used in the aforementioned studies), its existence is scheme-independent.

\begin{figure}[t]
\vspace{0.25cm}
\begin{minipage}[b]{0.5\linewidth}
\centering
\includegraphics[scale=0.35]{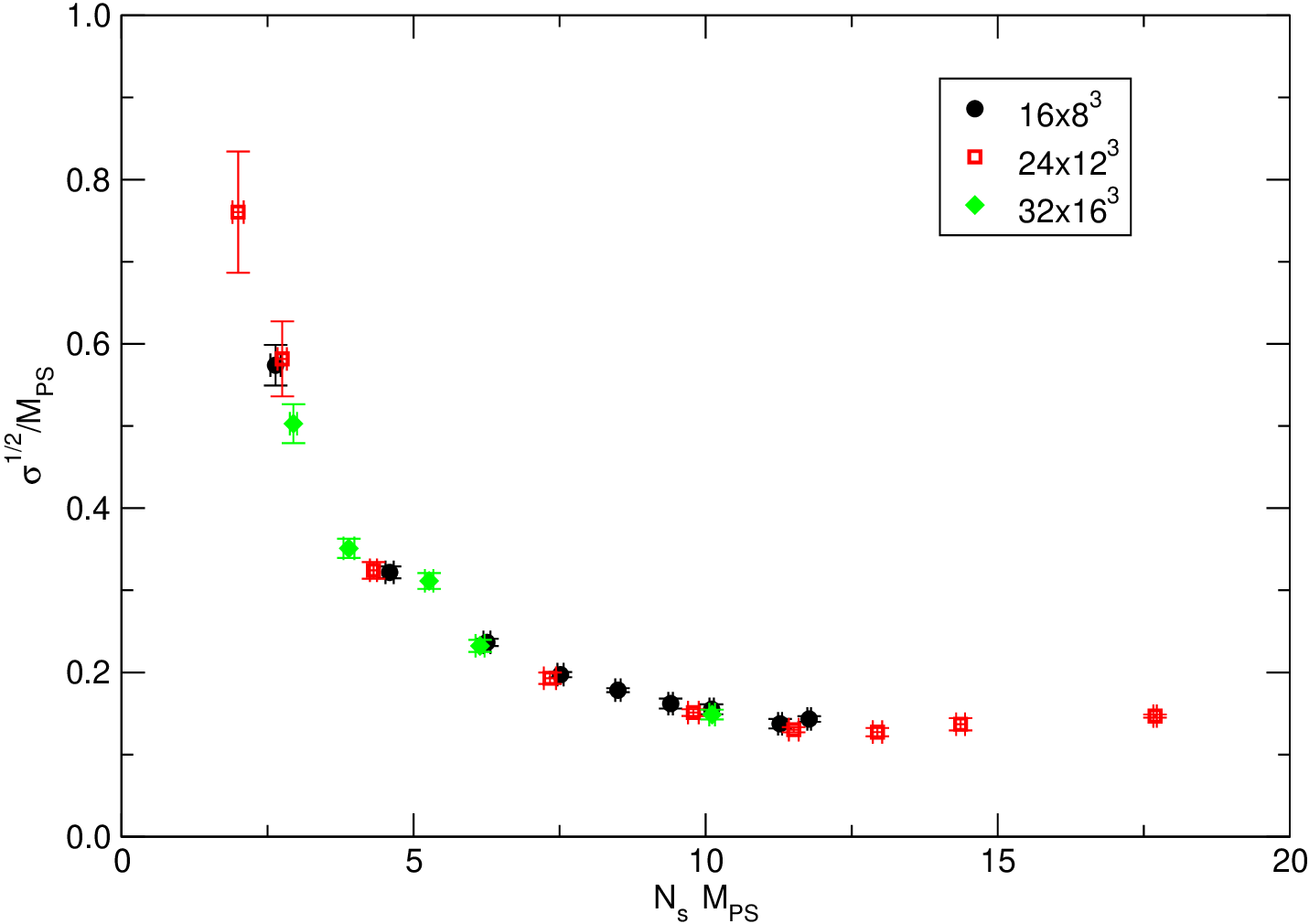}
\caption{The ratio $\sqrt{\sigma}/M_{PS}$ vs. $M_{PS} N_s$: data at various $N_s$ follow a unique curve, in agreement with finite size scaling predictions.}
\label{fig:4}
\end{minipage}
\hspace{0.1cm}
\begin{minipage}[b]{0.5\linewidth}
\centering
\includegraphics[scale=0.35]{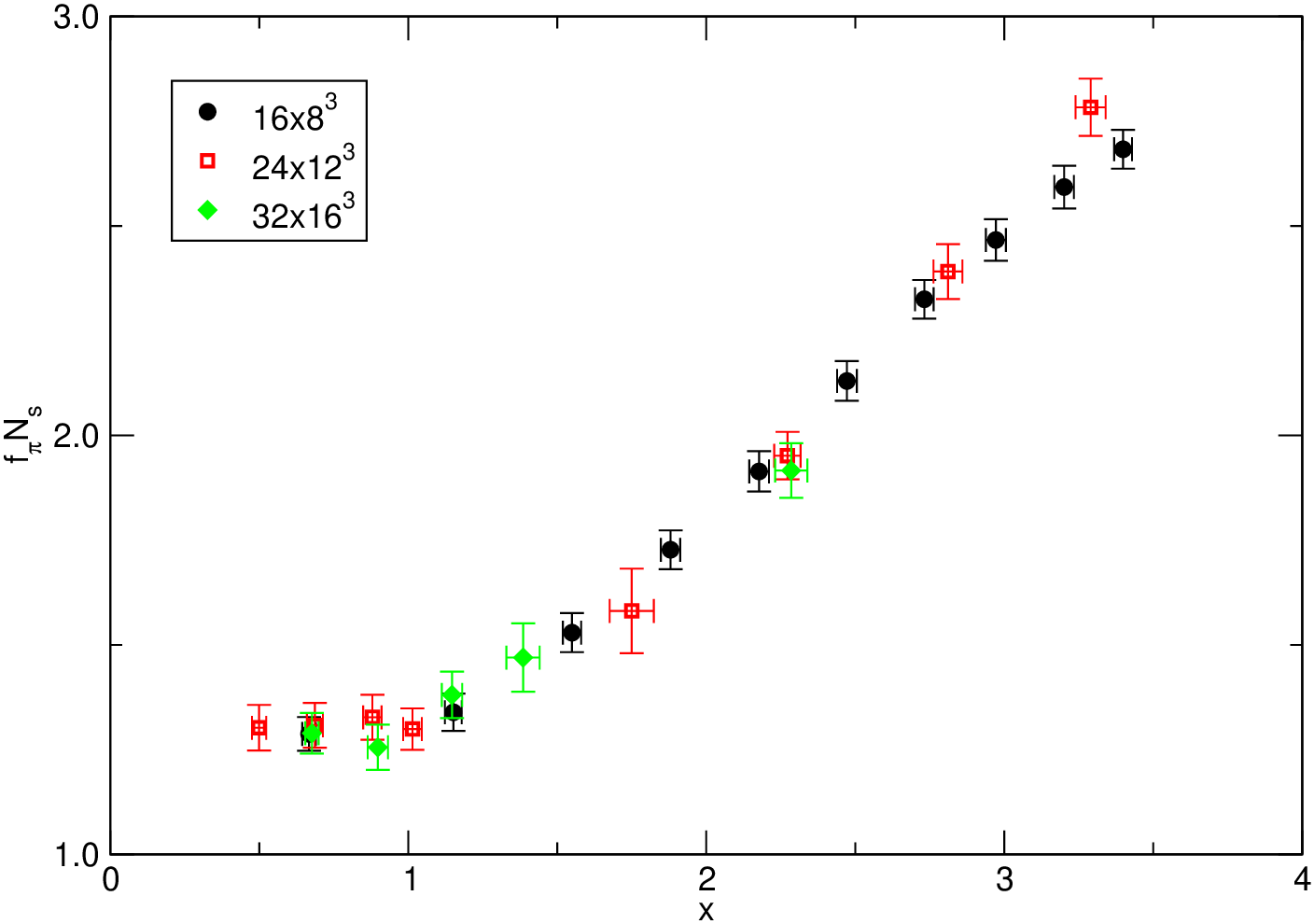}
\caption{Quality of the scaling for the pseudoscalar decay constant $f_{\pi}$ as a function of $x = N_s m^{1/(1+\gamma)}$ for $\gamma = 0.1$.}
\label{fig:5}
\end{minipage}
\end{figure}

When decreasing $m$ towards zero in theories with an IR fixed point, at some value of $m$ the correlation length will become of the order of the lattice size, the spectrum (and in particular the smallest mass in the spectrum) going to zero as a power of $m_{PCAC}$. When the correlation length $\xi$ becomes of the order of $N_s$ the observables display strong finite size effects. When these effects start to appear, the analysis described above can not be performed, which limits the range of PCAC masses near zero that is accessible once the spatial size is fixed to $N_s$. However, finite size effects can be used to characterise the approach to the zero mass limit using techniques of finite size scaling, in the same spirit in which second order phase transitions can be studied on finite systems near the critical point in Statistical Mechanics. This approach allows a direct access to the exponent governing the collapse of the spectrum to zero (we recall that for a mass $M$ in the spectrum in the chiral regime $M \propto m^{1/(1 + \gamma)}$, with $\gamma$ the anomalous dimension of the condensate). Finite size scaling predicts $M N_s = G(x)$, where $x = N_s m^{1/(1 + \gamma)}$ and $G$ is an unknown function. Hence, the rescaled quantity $M N_s$ is a universal function of $ N_s m^{1/(1 + \gamma)}$. This can be checked by considering the ratio of two quantities with mass dimension one $M_2/M_1$ as a function of e.g. $N_s M_2$. Fig.~\ref{fig:4} shows the ratio $M_{PS}/\sqrt{\sigma}$ vs. $N_s M_{PS}$: the collapse of the points onto a universal curve is striking. In order to determine the exponent $\gamma$, the scaling analysis need to be performed in terms of $m_{PCAC}$. \cite{Pica:2009hc} have shown that $f_{\pi}$ has strong finite size effects (see Fig. 3 on their article). Fig.~\ref{fig:5} shows that the curves at various $N_s$ collapse onto a single curve when $f_{\pi} N_s$ is plotted against $x$ with $\gamma = 0.1$. From our data, we conclude $0.05 \le \gamma \le 0.25$. This range of values of $\gamma$ automatically excludes the scaling $f_{\pi} \propto \gamma/(1 + \gamma)$ advocated by \cite{Sannino:2008pz} for $\gamma < 1$. We have performed also an analysis looking at the power behaviour of the string tension as the mass goes to zero; from this analysis, we get $0.1 \le \gamma \le 0.3$, which is broadly compatible with the determination obtained by studying $f_{\pi}$. Another independent determination comes from the analysis performed in the Lattice Schr\"odinger Functional formalism in~\cite{Bursa:2009we}, who finds $0.05 \le \gamma \le 0.56$, which is the most conservative estimate for $\gamma$. These values are smaller than the phenomenologically acceptable values ($\gamma \geq 1$). Hence, although Minimal Walking Technicolour fulfils the request of (near-)conformality, the anomalous dimension of the condensate seems to exclude its viability as a candidate of a beyond the Standard Model mechanism of Electroweak symmetry breaking in the canonical walking scenario (see e.g.~\cite{Evans:2005pu} for a mechanism compatible with a small $\gamma$).\\
\section*{Conclusions}
If taken by itself, none of the evidences presented in the previous section is a proof of the existence of an IR fixed point in Minimal Walking Technicolour, one noticeable limitation being that the simulations are performed at one single value of $\beta$ and hence there is no control over the extrapolation to the continuum limit. Moreover, it is not possible to exclude at this stage that those features do not survive at smaller fermion masses. However, various independent works that use different techniques all point to the existence of an IR fixed point for this theory. Nevertheless, to establish this without any doubt, better control over the continuum and the chiral limits need to be gained. It is worth remarking that while current simulations suggest conformal dynamics in the infrared, it can not be excluded that at higher distances than those investigated the theory become confining, i.e. that the theory be nearly-conformal in the infrared.

The first determinations of the anomalous dimension of the chiral condensate are starting to appear, and the general understanding is that this quantity is small: $0.05 \le \gamma \le 0.56$ is a conservative estimate. In order to be viable for phenomenology, Minimal Walking Technicolour should have an anomalous dimension of order one or bigger. Hence, those studies seem to exclude the relevance of the theory as an extension of the Standard Model. It remains to be seen whether other representations and higher values of $N$ have a higher $\gamma$. Recent results for SU(3) with sextet fermions point towards $\gamma \simeq 0.5$ (\cite{DeGrand:2009hu}). This might indicate that for fermions in the symmetric representation (which for SU(2) is equivalent to the adjoint representation) the anomalous dimension increases with the number of colours. If this is the case, a phenomenological viable candidate for strong dynamics beyond the Standard Model would have a gauge group SU($N$) with $N$ larger than three. It would be interesting to have a comparison with results for the anomalous dimension obtained in the case of fermions in the fundamental and in the adjoint representation and number of flavours close to the onset of the conformal window. One crucial concept that has been pointed out in~\cite{DelDebbio:2009fd} and elaborated further in this work is that the investigation of the gluonic part of the spectrum (i.e. the values of the string tension and of glueball masses) in connection with mesonic quantities can play a fundamental role in characterising the phenomenology of strong interactions beyond the Standard Model.

Ultimately, the experiments will unveil what is the fundamental theory that gives rise to the Standard Model phenomenology. If this is a strongly interacting field theory, with or without supersymmetry, the Lattice will have a crucial role for characterising its signature.
\section*{Acknowledgements}
This work draws heavily on results obtained in collaboration with L. Del Debbio, A. Patella, C. Pica and A. Rago. I am indebted with A. Patella for his useful comments on this manuscript. Discussions with M. Piai on various aspects of Technicolour and correspondence with F. Sannino are also grateful acknowledged. My work is supported by the Royal Society through the University Research Fellowship scheme.  

\bibliographystyle{rspublicnat}
\bibliography{bsmlattice}

\begin{thebibliography}{31}
\providecommand{\natexlab}[1]{#1}
\expandafter\ifx\csname urlstyle\endcsname\relax
  \providecommand{\doi}[1]{doi:\discretionary{}{}{}#1}\else
  \providecommand{\doi}{doi:\discretionary{}{}{}\begingroup
  \urlstyle{rm}\Url}\fi

\bibitem[{Accomando \emph{et~al.}(2006)}]{Accomando:2006ga}
Accomando, E. \emph{et~al.} 2006 {Workshop on CP Studies and Non-Standard Higgs
  Physics}.

\bibitem[{Amsler \emph{et~al.}(2008)}]{Amsler:2008zzb}
Amsler, C. \emph{et~al.} 2008 {Review of particle physics}.
\newblock \emph{Phys. Lett.}, \textbf{B667}, 1.
\newblock (\doi{10.1016/j.physletb.2008.07.018})

\bibitem[{Appelquist \emph{et~al.}(1986)Appelquist, Karabali \&
  Wijewardhana}]{Appelquist:1986an}
Appelquist, T.~W., Karabali, D. \& Wijewardhana, L. C.~R. 1986 {Chiral
  Hierarchies and the Flavor Changing Neutral Current Problem in Technicolor}.
\newblock \emph{Phys. Rev. Lett.}, \textbf{57}, 957.
\newblock (\doi{10.1103/PhysRevLett.57.957})

\bibitem[{Armoni(2009)}]{Armoni:2009jn}
Armoni, A. 2009 {The Conformal Window from the Worldline Formalism}.

\bibitem[{Bursa \emph{et~al.}(2009)Bursa, Del~Debbio, Keegan, Pica \&
  Pickup}]{Bursa:2009we}
Bursa, F., Del~Debbio, L., Keegan, L., Pica, C. \& Pickup, T. 2009 {Mass
  anomalous dimension in SU(2) with two adjoint fermions}.

\bibitem[{Catterall \emph{et~al.}(2008)Catterall, Giedt, Sannino \&
  Schneible}]{Catterall:2008qk}
Catterall, S., Giedt, J., Sannino, F. \& Schneible, J. 2008 {Phase diagram of
  SU(2) with 2 flavors of dynamical adjoint quarks}.
\newblock \emph{JHEP}, \textbf{11}, 009.
\newblock (\doi{10.1088/1126-6708/2008/11/009})

\bibitem[{Catterall \emph{et~al.}(2009)Catterall, Giedt, Sannino \&
  Schneible}]{Catterall:2009sb}
Catterall, S., Giedt, J., Sannino, F. \& Schneible, J. 2009 {Probes of nearly
  conformal behavior in lattice simulations of minimal walking technicolor}.

\bibitem[{Catterall \& Sannino(2007)}]{Catterall:2007yx}
Catterall, S. \& Sannino, F. 2007 {Minimal walking on the lattice}.
\newblock \emph{Phys. Rev.}, \textbf{D76}, 034\,504.
\newblock (\doi{10.1103/PhysRevD.76.034504})

\bibitem[{DeGrand(2009)}]{DeGrand:2009hu}
DeGrand, T. 2009 {Finite-size scaling tests for SU(3) lattice gauge theory with
  color sextet fermions}.

\bibitem[{Del~Debbio \emph{et~al.}(2009)Del~Debbio, Lucini, Patella, Pica \&
  Rago}]{DelDebbio:2009fd}
Del~Debbio, L., Lucini, B., Patella, A., Pica, C. \& Rago, A. 2009 {Conformal
  vs confining scenario in SU(2) with adjoint fermions}.

\bibitem[{Del~Debbio \emph{et~al.}(2008)Del~Debbio, Patella \&
  Pica}]{DelDebbio:2008zf}
Del~Debbio, L., Patella, A. \& Pica, C. 2008 {Higher representations on the
  lattice: numerical simulations. SU(2) with adjoint fermions}.

\bibitem[{Dietrich \& Sannino(2007)}]{Dietrich:2006cm}
Dietrich, D.~D. \& Sannino, F. 2007 {Conformal window of SU(N) gauge theories
  with fermions in higher dimensional representation}.
\newblock \emph{Phys. Rev.}, \textbf{D75}, 085\,018.
\newblock (\doi{10.1103/PhysRevD.75.085018})

\bibitem[{Dimopoulos \& Susskind(1979)}]{Dimopoulos:1979es}
Dimopoulos, S. \& Susskind, L. 1979 {Mass Without Scalars}.
\newblock \emph{Nucl. Phys.}, \textbf{B155}, 237--252.
\newblock (\doi{10.1016/0550-3213(79)90364-X})

\bibitem[{Eichten \& Lane(1980)}]{Eichten:1979ah}
Eichten, E. \& Lane, K.~D. 1980 {Dynamical Breaking of Weak Interaction
  Symmetries}.
\newblock \emph{Phys. Lett.}, \textbf{B90}, 125--130.
\newblock (\doi{10.1016/0370-2693(80)90065-9})

\bibitem[{Evans \& Sannino(2005)}]{Evans:2005pu}
Evans, N. \& Sannino, F. 2005 {Minimal walking technicolour, the top mass and
  precision electroweak measurements}.

\bibitem[{Fleming(2008)}]{Fleming:2008gy}
Fleming, G.~T. 2008 {Strong Interactions for the LHC}.
\newblock \emph{PoS}, \textbf{LATTICE2008}, 021.

\bibitem[{Hietanen \emph{et~al.}(2009{\natexlab{\emph{a}}})Hietanen,
  Rantaharju, Rummukainen \& Tuominen}]{Hietanen:2008mr}
Hietanen, A.~J., Rantaharju, J., Rummukainen, K. \& Tuominen, K.
  2009{\natexlab{\emph{a}}} {Spectrum of SU(2) lattice gauge theory with two
  adjoint Dirac flavours}.
\newblock \emph{JHEP}, \textbf{05}, 025.
\newblock (\doi{10.1088/1126-6708/2009/05/025})

\bibitem[{Hietanen \emph{et~al.}(2009{\natexlab{\emph{b}}})Hietanen,
  Rummukainen \& Tuominen}]{Hietanen:2009az}
Hietanen, A.~J., Rummukainen, K. \& Tuominen, K. 2009{\natexlab{\emph{b}}}
  {Evolution of the coupling constant in SU(2) lattice gauge theory with two
  adjoint fermions}.

\bibitem[{Hill \& Simmons(2003)}]{Hill:2002ap}
Hill, C.~T. \& Simmons, E.~H. 2003 {Strong dynamics and electroweak symmetry
  breaking}.
\newblock \emph{Phys. Rept.}, \textbf{381}, 235--402.
\newblock (\doi{10.1016/S0370-1573(03)00140-6})

\bibitem[{Holdom(1985)}]{Holdom:1984sk}
Holdom, B. 1985 {Techniodor}.
\newblock \emph{Phys. Lett.}, \textbf{B150}, 301.
\newblock (\doi{10.1016/0370-2693(85)91015-9})

\bibitem[{Miransky(1999)}]{Miransky:1998dh}
Miransky, V.~A. 1999 {Dynamics in the conformal window in {QCD} like theories}.
\newblock \emph{Phys. Rev.}, \textbf{D59}, 105\,003.
\newblock (\doi{10.1103/PhysRevD.59.105003})

\bibitem[{Pallante(2009)}]{Pallante}
Pallante, E. 2009 {Plenary talk given at Lattice 2009}.
\newblock \emph{PoS}, \textbf{LATTICE2009}.
\newblock To appear.

\bibitem[{Peskin \& Takeuchi(1992)}]{Peskin:1991sw}
Peskin, M.~E. \& Takeuchi, T. 1992 {Estimation of oblique electroweak
  corrections}.
\newblock \emph{Phys. Rev.}, \textbf{D46}, 381--409.
\newblock (\doi{10.1103/PhysRevD.46.381})

\bibitem[{Pica \emph{et~al.}(2009)Pica, Del~Debbio, Lucini, Patella \&
  Rago}]{Pica:2009hc}
Pica, C., Del~Debbio, L., Lucini, B., Patella, A. \& Rago, A. 2009 {Technicolor
  on the Lattice}.

\bibitem[{Poppitz \& Unsal(2009)}]{Poppitz:2009tw}
Poppitz, E. \& Unsal, M. 2009 {Conformality or confinement (II): One-flavor
  CFTs and mixed-representation QCD}.

\bibitem[{Sannino(2009{\natexlab{\emph{a}}})}]{Sannino:2008pz}
Sannino, F. 2009{\natexlab{\emph{a}}} {Conformal Chiral Dynamics}.
\newblock \emph{Phys. Rev.}, \textbf{D80}, 017\,901.

\bibitem[{Sannino(2009{\natexlab{\emph{b}}})}]{Sannino:2009me}
Sannino, F. 2009{\natexlab{\emph{b}}} {Higher Representations Duals}.

\bibitem[{Sannino \& Tuominen(2005)}]{Sannino:2004qp}
Sannino, F. \& Tuominen, K. 2005 {Techniorientifold}.
\newblock \emph{Phys. Rev.}, \textbf{D71}, 051\,901.
\newblock (\doi{10.1103/PhysRevD.71.051901})

\bibitem[{Susskind(1979)}]{Susskind:1978ms}
Susskind, L. 1979 {Dynamics of Spontaneous Symmetry Breaking in the Weinberg-
  Salam Theory}.
\newblock \emph{Phys. Rev.}, \textbf{D20}, 2619--2625.
\newblock (\doi{10.1103/PhysRevD.20.2619})

\bibitem[{Weinberg(1976)}]{Weinberg:1975gm}
Weinberg, S. 1976 {Implications of Dynamical Symmetry Breaking}.
\newblock \emph{Phys. Rev.}, \textbf{D13}, 974--996.
\newblock (\doi{10.1103/PhysRevD.13.974})

\bibitem[{Yamawaki \emph{et~al.}(1986)Yamawaki, Bando \&
  Matumoto}]{Yamawaki:1985zg}
Yamawaki, K., Bando, M. \& Matumoto, K.-i. 1986 {Scale Invariant Technicolor
  Model and a Technidilaton}.
\newblock \emph{Phys. Rev. Lett.}, \textbf{56}, 1335.
\newblock (\doi{10.1103/PhysRevLett.56.1335})

\end{thebibliography}
\end{document}